\documentclass[]{spie}  

\usepackage{amsmath,amsfonts,amssymb}
\usepackage{graphicx}
\usepackage[colorlinks=true, allcolors=blue]{hyperref}
\usepackage{aas_macros}

\title{The Roman Coronagraph Community Participation Program: calibration strategy for the Mueller matrix using on-sky sources}

\author[a,b]{Toshiyuki Mizuki}

\author[c]{Justin Hom}
\author[d]{Bertrand Mennesson}
\author[c]{Ramya M. Anche}
\author[e]{Maxwell A. Millar-Blanchaer}
\author[d]{Vanessa P. Bailey}

\author[f]{Koji Kawabata}
\author[f]{Mitsuhiko Takeuchi}
\author[g]{Motohide Tamura}
\author[a,b]{Naoshi Murakami}

\author[a,b]{John Livingston}
\author[h,i]{Jason J. Wang}
\author[c]{Schuyler G. Wolff}
\author[j]{Guillermo Gonzalez}
\author[e]{Eric Shen}
\author[k]{Tsutsumi Nagai}
\author[a,b]{Taichi Uyama}

\author{the Roman Coronagraph Community Participation Program Team}

\affil[a]{Astrobiology Center, 2-21-1 Osawa, Mitaka, Tokyo 181-8588, Japan}
\affil[b]{National Astronomical Observatory of Japan, 2-21-1 Osawa, Mitaka, Tokyo 181-8588, Japan}
\affil[c]{Steward Observatory and the Department of Astronomy, The University of Arizona, 933 N Cherry Ave, Tucson, AZ, 85721, USA}
\affil[d]{Jet Propulsion Laboratory, California Institute of Technology, 4800 Oak Grove Drive, Pasadena, CA 91109, USA}
\affil[e]{University of California, Santa Barbara, CA 93106, USA}
\affil[f]{Hiroshima Astrophysical Science Center, Hiroshima University, Higashi-Hiroshima, Hiroshima 739-8526, Japan}
\affil[g]{Department of Astronomy, Graduate School of Science, The University of Tokyo, 7-3-1, Hongo, Bunkyo-ku, Tokyo 113-0033, Japan}
\affil[h]{Northwestern University, 633 Clark Street Evanston, IL 60208, USA}
\affil[i]{Center for Interdisciplinary Exploration and Research in Astrophysics, 1800 Sherman Ave, Evanston, IL 60201, USA}
\affil[j]{Tellus1 Scientific, LLC, 8401 Whitesburg Dr SE, Unit 4662, Huntsville, AL 35802 USA}
\affil[k]{Department of Astronomical Science, The Graduate University for Advanced Studies, SOKENDAI, 2-21-1 Osawa, Mitaka, Tokyo 181-8588, Japan}

\authorinfo{Further author information: (Send correspondence to Toshiyuki Mizuki)\\Toshiyuki Mizuki: E-mail: toshiyuki.mizuki.astr@gmail.com}

\begin{document} 
\maketitle

\newcommand{\cgi}{Roman Coronagraph Instrument}

\begin{abstract}
The Nancy Grace Roman Space Telescope Coronagraph Instrument will provide space-based polarimetric observations of circumstellar disks and exoplanetary systems. Accurate reconstruction of the linear polarization fraction requires calibration of the instrumental Mueller matrix using polarized and weakly polarized standard stars. We constructed a candidate catalog by combining published optical polarimetry with Gaia DR3 astrometry and photometry and selected separate samples for coronagraphic calibration observations and observations using a neutral-density filter. Precursor $VRI$-band polarimetry of 18 faint candidates was obtained with HONIR on the 1.5-m Kanata telescope. The wavelength dependence of their normalized Stokes parameters was modeled using the Serkowski law to predict their polarization properties in \cgi\ Bands~1 and 4, and 2 sets of 3 calibrators were selected for the 2 calibration scenarios. We then estimated the achievable LPF reconstruction accuracy using Monte Carlo simulations that include uncertainties in the calibrator polarization properties, photometric noise, and residual detector-response errors. A dithered observing configuration was also simulated to reduce differential detector-response errors among the calibrators. The current estimates indicate LPF reconstruction errors at the few-percentage-point level, with a small bias arising from treating a weakly polarized calibrator as unpolarized. Finally, we present progress toward an end-to-end test using \texttt{corgisim} and \texttt{corgiDRP}, including successful processing of simulated datasets from Level~1 through Level~2b.
\end{abstract}


\section{INTRODUCTION}
The Nancy Grace Roman Space Telescope is NASA's next flagship astrophysics mission and is scheduled for launch on August 30, 2026. The \cgi\ is a technology demonstrator for space-based high-contrast imaging with active wavefront control and is allocated a 90-day observing phase within the first 18 months of the mission. Its formal performance requirement is to measure, with a signal-to-noise ratio of at least 5, a point source with a flux ratio of at least $10^{-7}$ at angular separations of $6$--$9\lambda/D$ from a star as faint as $V_{\rm AB}=5$, using a bandpass centered below 600~nm with a fractional bandwidth of at least $10\%$. Beyond this threshold requirement, the \cgi\ is expected to achieve contrast levels of approximately $10^{-8}$ or better in some observing configurations (Mennesson et al., 2022)\cite{2022SPIE12180E..1WM}. Potential observations include debris disks, self-luminous exoplanets, and reflected-light observations of planets detected through radial-velocity or astrometric measurements. In addition to the formally required Band~1 imaging mode, the \cgi\ includes polarimetric and spectroscopic capabilities that will be supported on a best-effort basis (Groff et al., 2025)\cite{2025JATIS..11c1510G}.

High-contrast polarimetry provides unique constraints on the physical properties of planetary systems. For debris disks, the polarization fraction and its spatial distribution depend on both the scattering geometry and the properties of the dust grains, enabling constraints on disk geometry, the scattering phase function, and grain size, composition, and porosity (e.g., Tazaki$\&$Dominik, 2022)\cite{2022A&A...663A..57T}. Observationally, these effects can produce striking differences between disk morphologies seen in total and polarized intensity, providing complementary views of the same disk structure (e.g., Perrin et al., 2015)\cite{2015ApJ...799..182P}. Polarimetry also has the potential to characterize the atmospheres of reflected-light exoplanets because the degree and wavelength dependence of polarization are sensitive to molecular scattering and to the composition and vertical distribution of clouds and hazes (Stam et al., 2004)\cite{2004A&A...428..663S}. From an instrumental perspective, polarization aberrations caused by differential reflection from telescope mirrors generate polarized speckles that can limit both polarimetric accuracy and achievable coronagraphic contrast (Doelman et al., 2023)\cite{2023SPIE12680E..0VD}. Characterizing these effects with the \cgi\ will also provide important lessons for the design and performance evaluation of the Habitable Worlds Observatory, where polarization aberrations may become a critical limitation at contrast levels approaching $10^{-10}$.

Polarimetry with the \cgi\ uses two Wollaston prisms, WOL0 and WOL45, whose transmission axes are offset by $45^\circ$. Switching between the two prisms provides the complementary measurements required to derive Stokes $Q$ and $U$. Unlike many ground-based polarimeters, the \cgi\ does not include a rotating polarization modulator, such as a half-wave plate, thereby reducing the number of moving optical components. Nevertheless, instrumental polarization and cross-talk introduced by the optical system must be characterized through the Mueller matrix. Because circular polarization is assumed to be negligible and is not measured by the \cgi, a reduced $3\times3$ Mueller matrix is used to relate the incident and observed Stokes vectors, $(I,Q,U)$. The current model Mueller matrices were derived from end-to-end optical modeling conducted by the Roman Coronagraph project team\footnote{\url{https://roman.ipac.caltech.edu/docs/Roman-Coronagraph-Optical-Model-Mueller-Matrices-450-to-950nm.pdf}}. The predicted matrices are nearly diagonal in both Band~1 and Band~4, with $I$-to-$Q$ leakage corresponding to an apparent polarization of approximately $0.9\%$ and $0.4\%$, respectively, for an intrinsically unpolarized source. The model also predicts negligible $Q$--$U$ cross-talk and a lower response to Stokes $U$ in Band~4 than to $I$ and $Q$. Because the \cgi\ does not include an onboard polarized calibration source that provides an absolute polarization reference, the Mueller-matrix coefficients must be calibrated using observations of polarized and weakly polarized standard stars.

\section{Selection of Polarimetric Standard Stars}
Polarimetric calibration of the \cgi\ requires standard stars that satisfy both the instrumental observability constraints and the requirements for accurate polarization calibration. We constructed a candidate catalog by combining published optical polarimetry with astrometric and photometric information from Gaia DR3 (Gaia Collaboration et al., 2023)\cite{2023A&A...674A...1G}. The initial sample was drawn primarily from the optical polarization catalog compiled by Panopoulou et al. (2025)\cite{Panopoulou25} and was supplemented with measurements from Whittet et al. (1992)\cite{1992ApJ...386..562W} and Blinov et al. (2023)\cite{2023A&A...677A.144B}. The additional catalogs were cross-matched with Gaia DR3, and duplicate entries were removed. We restricted the sample to high ecliptic latitudes, $|\beta|>54^\circ$, approximately corresponding to the Roman continuous viewing zones, and retained sources with Gaia $G\leq14$~mag to enable efficient observations at high signal-to-noise ratio. To minimize contamination from nearby sources and ensure reliable target acquisition, we rejected stars with a neighboring Gaia source within $12''$ and $\Delta G<3$~mag. We also required ${\rm RUWE}\leq1.4$ to exclude likely unresolved binaries and sources with anomalous astrometric solutions.

We considered separate faint and bright samples for the 2 polarimetric-calibration scenarios because the ND filter may introduce its own instrumental polarization and cross-talk. Faint standards are required to characterize the Mueller matrix of the instrument without the ND filter, matching the optical configuration used for coronagraphic science observations. Stars brighter than approximately $V=10.9$~mag are expected to saturate in this configuration. We therefore applied an initial criterion of $G\gtrsim10.9$~mag to the faint sample and subsequently examined the available $V$-band photometry for candidates near this cutoff, requiring $V\gtrsim10.9$~mag. Because many stars in the faint sample have only a single published polarization measurement, the precursor observations were designed to obtain additional multi-band measurements and estimate their polarization in \cgi\ Bands~1 and 4 using the Serkowski law (Serkowski et al., 1975)\cite{1975ApJ...196..261S}.
Bright standards ($G<9$~mag) with multiple ground-based polarization measurements were retained as candidates for observations with the ND filter, allowing the polarization response of the combined instrument and ND-filter configuration to be calibrated separately.

We selected both polarized and weakly polarized standard stars. The estimate of the achievable linear polarization fraction (LPF) accuracy by Zellem et al. (2022)\cite{2022SPIE12180E..1ZZ} assumes polarized calibrators with intrinsic polarization fractions of a few percent. We therefore adopted $p\geq2\%$ for polarized standards to ensure sufficiently strong signals for constraining the instrumental response. For weakly polarized standards, we adopted $p\leq0.5\%$, while retaining their measured polarization as a nonzero calibration input. After applying these criteria, 8 bright and 338 faint stars remained as candidate polarimetric standards for the \cgi.

\section{Precursor Observations with HONIR}
We observed 18 of the 338 candidate standards that were observable between October 10, 2025, and January 14, 2026, using HONIR (Hiroshima Optical and Near-InfraRed camera; Akitaya et al., 2014)\cite{2014SPIE.9147E..4OA}, mounted on the 1.5-m Kanata telescope at Higashi-Hiroshima Observatory. HONIR is equipped with a rotating achromatic half-wave plate and a fixed Wollaston prism. For each target, the exposure time per frame was adjusted according to its brightness. Each observing set consisted of 4 frames obtained at half-wave-plate position angles of $0^\circ$, $45^\circ$, $22.5^\circ$, and $67.5^\circ$. Three such sets were acquired in each of the $V$, $R$, and $I$ bands.
The data were reduced using the HONIR pipeline (Nakamura 2022; Hori 2023)\cite{Nakamura22,Hori23}, which performs standard preprocessing and polarimetric calibration. From the reduced data, we derived the normalized Stokes parameters $q$ and $u$, together with the polarization fraction and polarization position angle.

\section{Serkowski-Law Fitting for Roman Bands~1 and 4}
We modeled the wavelength dependence of the linear polarization in Stokes space. For each measurement, the observed polarization fraction $p$ and electric-vector position angle (EVPA, $\theta$) were converted into the normalized Stokes parameters,
\begin{equation}
q=p\cos 2\theta,
\qquad
u=p\sin 2\theta.
\end{equation}
We then fitted $q(\lambda)$ and $u(\lambda)$ simultaneously, assuming a common wavelength dependence described by the Serkowski law (Serkowski et al., 1975)\cite{1975ApJ...196..261S},
\begin{equation}
S(\lambda)
=
\exp\left[
-K\ln^2\left(\frac{\lambda_{\max}}{\lambda}\right)
\right],
\end{equation}
such that
\begin{equation}
q(\lambda)=q_{\max}S(\lambda),
\qquad
u(\lambda)=u_{\max}S(\lambda).
\end{equation}
Here, $q_{\max}$ and $u_{\max}$ are independent amplitudes, whereas $\lambda_{\max}$ and $K$ are shared by the 2 Stokes components. This formulation assumes that the polarization position angle is independent of wavelength, while the polarization amplitude follows the Serkowski law. The fitted maximum polarization fraction and position angle are given by
\begin{equation}
p_{\max}
=
\sqrt{q_{\max}^{2}+u_{\max}^{2}},
\qquad
\theta
=
\frac{1}{2}\operatorname{atan2}
\left(u_{\max},q_{\max}\right),
\end{equation}
where the 2-argument arctangent ensures that the correct quadrant is selected.
We adopted the original Serkowski-law value of $K=1.15$, which produced smaller residuals in both $q(\lambda)$ and $u(\lambda)$ than the alternative wavelength-dependent prescriptions examined in this work. For weakly polarized stars, for which $\lambda_{\max}$ could not be meaningfully constrained, we fixed $\lambda_{\max}=550$~nm. For the remaining targets, $\lambda_{\max}$ was fitted simultaneously with $q_{\max}$ and $u_{\max}$.

For each observed band, the measurements were compared with the band-averaged model Stokes parameters. The model value of each Stokes parameter in a given bandpass was calculated as
\begin{equation}
x_{\rm band}
=
\frac{
\int x(\lambda)F_{\star}(\lambda)R(\lambda),d\lambda
}{
\int F_{\star}(\lambda)R(\lambda),d\lambda
},
\qquad x\in{q,u},
\end{equation}
and the corresponding polarization fraction was calculated as
\begin{equation}
p_{\rm band}
=
\sqrt{q_{\rm band}^{2}+u_{\rm band}^{2}},
\end{equation}
where $F_{\star}(\lambda)$ is the stellar spectrum and $R(\lambda)$ is the filter-response function. The passbands of the published measurements were approximated using the corresponding filter-response curves from the \texttt{pyphot} filter library. For each target, $F_{\star}(\lambda)$ was approximated using a Kurucz model atmosphere (Kurucz, 1993)\cite{kurucz93}. The effective temperature was estimated from the Gaia DR3 $G_{\rm BP}-G_{\rm RP}$ color using an approximate main-sequence relation based on the MESA Isochrones and Stellar Tracks (MIST; Dotter, 2016)\cite{mist1}. We adopted $\log g=4.5$ and $[{\rm Fe/H}]=0.0$ for all stars.

The fit was performed by least-squares minimization, and the parameter uncertainties were estimated from the covariance matrix. When the reduced chi-square,
\begin{equation}
\chi_{\nu}^{2}
=
\frac{\chi^{2}}{\nu},
\end{equation}
exceeded unity, where $\nu$ is the number of degrees of freedom, the uncertainties were multiplied by an error-inflation factor,
\begin{equation}
f_{\rm inflate}
=
\sqrt{\chi_{\nu}^{2}}.
\end{equation}
The adopted parameter uncertainties were therefore calculated as
\begin{equation}
\sigma_{\rm adopted}
=
f_{\rm inflate}\sigma.
\end{equation}
No inflation was applied when $\chi_{\nu}^{2}\leq1$. This procedure empirically accounts for excess scatter among the measurements, including potential inter-survey systematics, intrinsic variability, and underestimated measurement uncertainties, while leaving the best-fit parameters unchanged.

\begin{figure*}[htbp]
\begin{center}
\includegraphics[width=\linewidth]{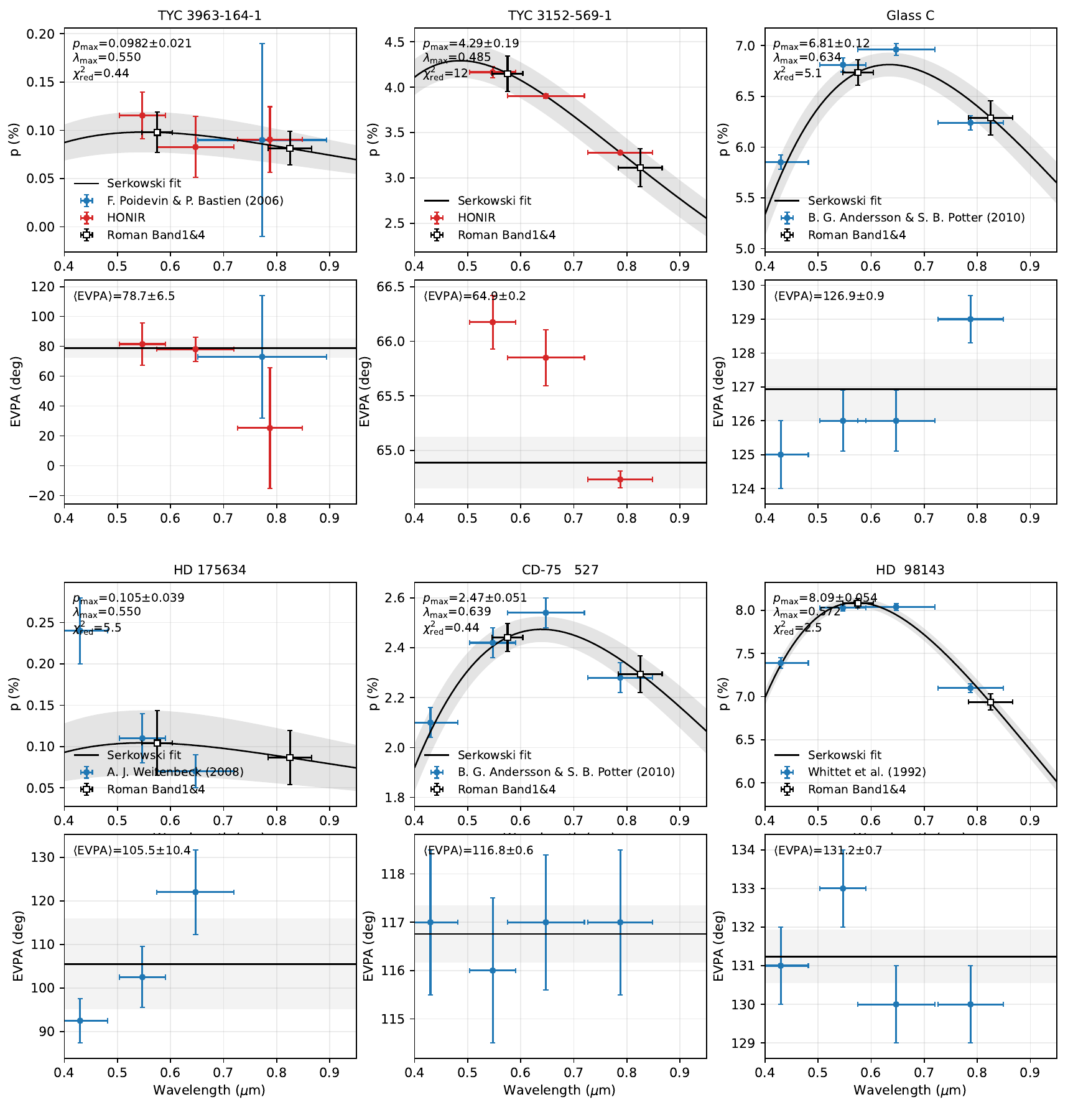}
\end{center}
\caption{Wavelength dependence of the polarization of the six polarimetric standards selected for \cgi\ calibration observations.}
\label{fig:serkowskifit}
\end{figure*}

Using the best-fit Serkowski parameters, we calculated the expected polarization fractions in \cgi\ Bands~1 and 4. We approximated Bands~1 and 4 as top-hat bandpasses centered at 575 and 825~nm, respectively, each with a fractional bandwidth of $10\%$. The expected polarization in each band was calculated using the same bandpass-averaging formalism, with $R(\lambda)$ replaced by the corresponding top-hat response. Uncertainties in the predicted polarization fractions were propagated from the inflated uncertainties in $p_{\max}$ and $\lambda_{\max}$. We evaluated the model at the $\pm1\sigma$ excursions of each parameter and added the resulting contributions in quadrature.
Table~\ref{table_commissionig} lists the 2 sets of 3 calibrators adopted for the 2 calibration scenarios planned for the early operation of \cgi. Each set consists of 1 nearly unpolarized standard and 2 polarized standards. Figure~\ref{fig:serkowskifit} shows the wavelength dependence of the polarization for the 6 standards selected for these observations.

\begin{table}[t]
\caption{Expected polarization properties of the selected standards in \cgi\ Bands~1 and 4.}
\label{table_commissionig}
\centering
\scriptsize
\begin{tabular}{cccccccccc}
\hline
Name & RA & Dec & $d$ & Main type & $V$ & $G_{\rm BP}-G_{\rm RP}$ & $p_{\rm B1}$ & $p_{\rm B4}$ & EVPA \\
 & (deg) & (deg) & (pc) &  & (mag) & (mag) & (\%) & (\%) & (deg) \\
\hline
TYC 3963-164-1 & 312.955 & 59.904 & 248.2 & Star & 10.98 & 0.82 & 0.098 $\pm$ 0.021 & 0.081 $\pm$ 0.017 & 78.7 $\pm$ 6.5 \\
TYC 3152-569-1 & 305.989 & 39.095 & 1242.8 & Star & 11.60 & 0.97 & 4.150 $\pm$ 0.196 & 3.111 $\pm$ 0.210 & 64.9 $\pm$ 0.2 \\
Glass C & 165.798 & -77.351 & 280.8 & TTauri* & 11.48 & 2.02 & 6.734 $\pm$ 0.126 & 6.288 $\pm$ 0.170 & 126.9 $\pm$ 0.9 \\
HD 175634 & 283.716 & 33.981 & 353.2 & Star & 7.68 & -0.07 & 0.104 $\pm$ 0.039 & 0.087 $\pm$ 0.033 & 105.5 $\pm$ 10.4 \\
CD-75   527 & 168.228 & -75.870 & 730.9 & Star & 9.14 & 1.60 & 2.441 $\pm$ 0.056 & 2.295 $\pm$ 0.074 & 116.8 $\pm$ 0.6 \\
HD  98143 & 168.820 & -77.518 & 268.4 & Star & 7.59 & 0.90 & 8.083 $\pm$ 0.054 & 6.938 $\pm$ 0.093 & 131.2 $\pm$ 0.7 \\
\hline
\end{tabular}
\end{table}

\section{Monte Carlo Estimate of LPF Reconstruction Accuracy}
\subsection{Monte Carlo Framework}

Following the framework of Zellem et al. (2022)\cite{2022SPIE12180E..1ZZ}, we estimated the achievable accuracy of the LPF reconstruction using a Monte Carlo simulation of the Mueller-matrix calibration. The instrumental polarization of the \cgi\ was represented by a wavelength-dependent $3\times3$ Mueller matrix acting on the Stokes parameters $(I,Q,U)$. Model Mueller matrices evaluated at 575 and 825~nm were adopted for Bands~1 and 4, respectively.

For calibrator $k$ observed at telescope roll $r$, the noiseless calibration Stokes vector adopted in the Mueller-matrix reconstruction is
\begin{equation}
{\bf S}_{{\rm cal},k,r}
=
I_k
\begin{pmatrix}
1 \\
p_{{\rm cal},k}\cos 2\theta'_{{\rm cal},k,r} \\
p_{{\rm cal},k}\sin 2\theta'_{{\rm cal},k,r}
\end{pmatrix},
\qquad
k=1,2,3,
\end{equation}
where the instrument-frame PA is
\begin{equation}
\theta'_{{\rm cal},k,r}
=
\theta_{{\rm cal},k}-\phi_r,
\end{equation}
and $\phi_r$ is the telescope roll angle.

In each Monte Carlo realization, the LPF and PA of each calibrator are perturbed according to their adopted uncertainties. The resulting simulated input Stokes vector is
\begin{equation}
{\bf S}_{{\rm sim},k,r}
=
I_k
\begin{pmatrix}
1 \\
p_{{\rm sim},k}\cos 2\theta'_{{\rm sim},k,r} \\
p_{{\rm sim},k}\sin 2\theta'_{{\rm sim},k,r}
\end{pmatrix},
\end{equation}
where
\begin{equation}
\theta'_{{\rm sim},k,r}
=
\theta_{{\rm sim},k}-\phi_r.
\end{equation}
Thus, ${\bf S}_{{\rm cal},k,r}$ represents the nominal calibrator properties assumed in the Mueller-matrix reconstruction, whereas ${\bf S}_{{\rm sim},k,r}$ includes uncertainties in those properties. For the weakly polarized standard, the simulated input is generated using its measured nonzero polarization, whereas the Mueller-matrix reconstruction assumes
\begin{equation}
p_{{\rm cal},1}=0.
\end{equation}

The simulated calibrator Stokes vectors are propagated through the true Mueller matrix. Residual detector-response errors and photometric noise are then applied, as described in the following subsection, to obtain the observed Stokes vectors,
\begin{equation}
{\bf S}_{{\rm obs},k,r}
=
\begin{pmatrix}
I_{{\rm obs},k,r} \\
Q_{{\rm obs},k,r} \\
U_{{\rm obs},k,r}
\end{pmatrix}.
\end{equation}

For each roll, the observed and calibration Stokes matrices are defined as
\begin{equation}
{\bf Y}_r
=
\begin{pmatrix}
{\bf S}_{{\rm obs},1,r} &
{\bf S}_{{\rm obs},2,r} &
{\bf S}_{{\rm obs},3,r}
\end{pmatrix}
\end{equation}
and
\begin{equation}
{\bf X}_r
=
\begin{pmatrix}
{\bf S}_{{\rm cal},1,r} &
{\bf S}_{{\rm cal},2,r} &
{\bf S}_{{\rm cal},3,r}
\end{pmatrix}.
\end{equation}
The 3 calibration Stokes vectors must be linearly independent for the Mueller matrix to be uniquely determined. The estimated Mueller matrix is
\begin{equation}
{\bf M}_{{\rm est},r}
=
{\bf Y}_r{\bf X}_r^{-1}.
\end{equation}
When the first standard is treated as exactly unpolarized, the estimated Mueller matrix is given explicitly by
\begin{equation}
{\bf M}_{{\rm est},r}
=
\begin{pmatrix}
I_{{\rm obs},1,r} & I_{{\rm obs},2,r} & I_{{\rm obs},3,r} \\
Q_{{\rm obs},1,r} & Q_{{\rm obs},2,r} & Q_{{\rm obs},3,r} \\
U_{{\rm obs},1,r} & U_{{\rm obs},2,r} & U_{{\rm obs},3,r}
\end{pmatrix}
\begin{pmatrix}
I_1 & I_2 & I_3 \\
0 &
I_2p_{{\rm cal},2}\cos 2\theta'_{{\rm cal},2,r} &
I_3p_{{\rm cal},3}\cos 2\theta'_{{\rm cal},3,r} \\
0 &
I_2p_{{\rm cal},2}\sin 2\theta'_{{\rm cal},2,r} &
I_3p_{{\rm cal},3}\sin 2\theta'_{{\rm cal},3,r}
\end{pmatrix}^{-1}.
\end{equation}

For the simulated target, the input Stokes vector at roll $r$ is
\begin{equation}
{\bf S}_{{\rm true},r}
=
I
\begin{pmatrix}
1 \\
p_{\rm true}\cos 2\theta'_{{\rm true},r} \\
p_{\rm true}\sin 2\theta'_{{\rm true},r}
\end{pmatrix},
\end{equation}
where
\begin{equation}
\theta'_{{\rm true},r}
=
\theta_{\rm true}-\phi_r.
\end{equation}
Residual detector-response errors and photometric noise are applied to the target in the same manner as for the calibrators, as described in the following subsection, to obtain the observed target Stokes vector,
\begin{equation}
{\bf S}_{{\rm obs},r}
=
\begin{pmatrix}
I_{{\rm obs},r} \\
Q_{{\rm obs},r} \\
U_{{\rm obs},r}
\end{pmatrix}.
\end{equation}
The target Stokes vector is reconstructed independently at each roll using
\begin{equation}
{\bf S}_{{\rm est},r}
=
{\bf M}_{{\rm est},r}^{-1}
{\bf S}_{{\rm obs},r}.
\end{equation}
The LPF reconstruction error is defined as
\begin{equation}
\Delta p
=
\frac{1}{2}
\sum_{r=1}^{2}
\frac{
\sqrt{Q_{{\rm est},r}^{2}+U_{{\rm est},r}^{2}}
}{
I_{{\rm est},r}
}
-
p_{\rm true}.
\end{equation}

We repeated this procedure over many MC realizations. The LPF bias and scatter were calculated as
\begin{equation}
b_p
=
\left\langle\Delta p\right\rangle
\end{equation}
and
\begin{equation}
\sigma_p
=
\sqrt{
\left\langle
\left(\Delta p-b_p\right)^2
\right\rangle
}.
\end{equation}
The total root-mean-square reconstruction error was calculated as
\begin{equation}
{\rm RMSE}_p
=
\sqrt{\left\langle\Delta p^2\right\rangle}
=
\sqrt{\sigma_p^2+b_p^2}.
\end{equation}
The bias, scatter, and RMSE are reported in percentage points.

\subsection{Detector-Response Model}

The residual detector-response errors are modeled in the 4 polarization-channel intensities corresponding to the nominal $0^\circ$, $90^\circ$, $45^\circ$, and $135^\circ$ analyzer outputs. For calibrator $k$ observed at telescope roll $r$, the observed Stokes vector is expressed as
\begin{equation}
{\bf S}_{{\rm obs},k,r}
=
{\bf T}
\left[
{\bf F}_{k,r}
{\bf A}
{\bf M}_{\rm true}(\lambda)
{\bf S}_{{\rm sim},k,r}
+
{\boldsymbol\epsilon}_{k,r}
\right],
\end{equation}
where ${\boldsymbol\epsilon}_{k,r}$ represents photometric noise in the 4 polarization channels. The matrix
\begin{equation}
{\bf A}
=
\frac{1}{2}
\begin{pmatrix}
1 & 1  & 0 \\
1 & -1 & 0 \\
1 & 0  & 1 \\
1 & 0  & -1
\end{pmatrix}
\end{equation}
converts the Stokes vector $(I,Q,U)$ into the 4 polarization-channel intensities, while
\begin{equation}
{\bf T}
=
\begin{pmatrix}
1/2 & 1/2 & 1/2 & 1/2 \\
1   & -1  & 0   & 0 \\
0   & 0   & 1   & -1
\end{pmatrix}
\end{equation}
converts the channel intensities back into the observed Stokes vector.
Under a multiplicative-response approximation, the residual flat-field and charge-transfer-inefficiency errors are represented by
\begin{equation}
{\bf F}_{k,r}
=
\begin{pmatrix}
f_{0,k,r} & 0 & 0 & 0 \\
0 & f_{90,k,r} & 0 & 0 \\
0 & 0 & f_{45,k,r} & 0 \\
0 & 0 & 0 & f_{135,k,r}
\end{pmatrix}.
\end{equation}

A key assumption of the calibration framework considered by Zellem et al. (2022)\cite{2022SPIE12180E..1ZZ} is that the 3 calibrators sample a common detector region. For detector samples shared by all 3 calibrators at a given roll, we assume
\begin{equation}
{\bf F}_{1,r}
=
{\bf F}_{2,r}
=
{\bf F}_{3,r}
\equiv
{\bf F}_r.
\end{equation}
The residual detector response is therefore treated as a correlated error shared by the 3 calibrators, rather than as an independent error for each calibrator. This common-region assumption reduces differential detector-response errors in the estimated Mueller matrix.

\subsection{Simulation of a Dithered Observing Strategy}

Because unocculted calibration observations rely on EXCAM acquisition, the calibrators cannot be guaranteed to fall at exactly the same detector position. In 2 demonstrated thermal-vacuum acquisition cases, the stellar image was placed at residual offsets of $0.050''$ and $0.027''$ from the target pixel after pointing correction (Fathpour et al., 2025)\cite{2025JATIS..11b1402F}. These results indicate that calibrator-dependent differences in detector position may occur and could introduce systematic errors through spatial variations in the detector response.

\begin{figure*}[htbp]
\begin{center}
\includegraphics[width=\linewidth]{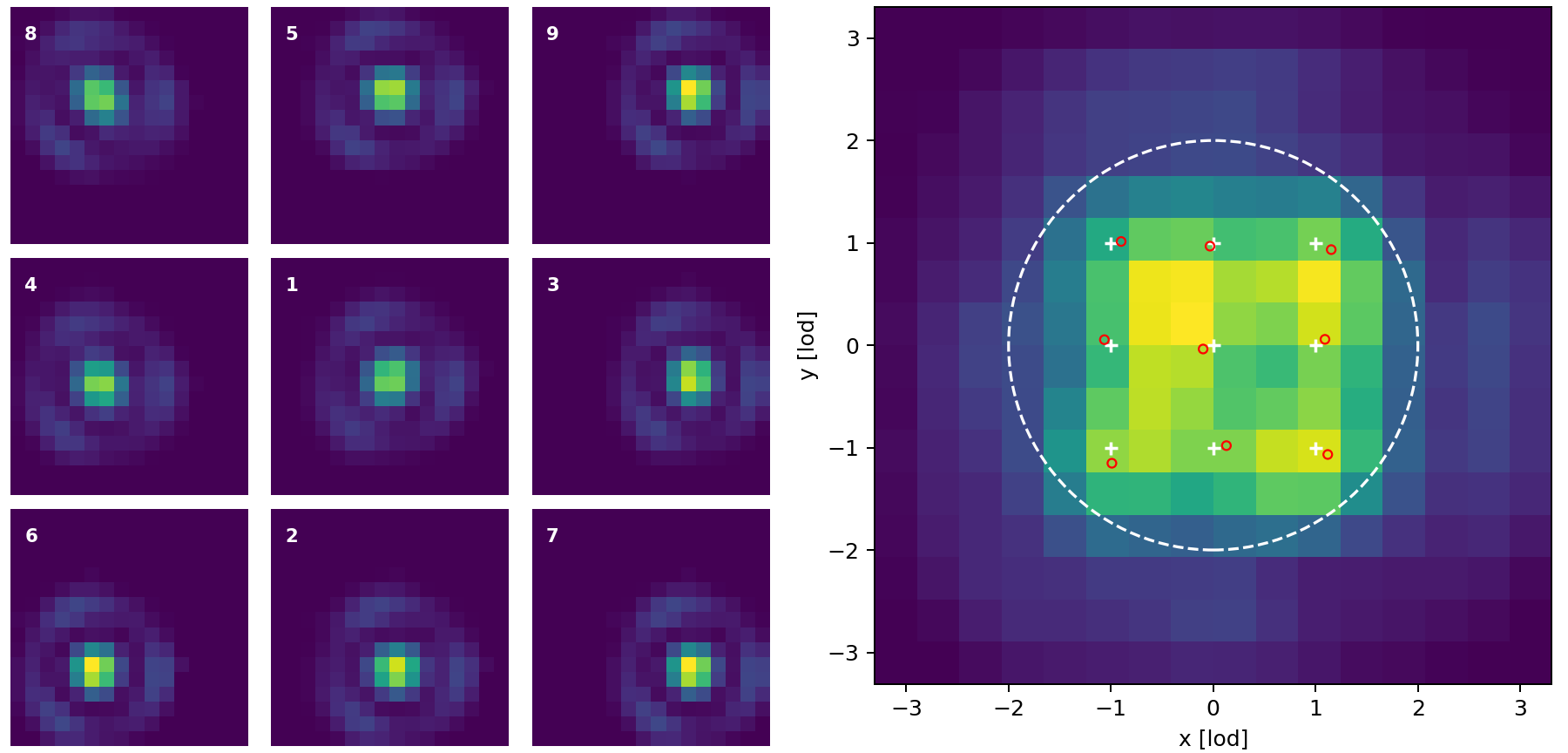}
\end{center}
\caption{
Left: Nine image slices in the data cube, obtained on a $3\times3$ dither grid with a spacing of $1\lambda/D$.
Right: The stacked image, assuming a relative positioning error of $0.1\lambda/D$ at each dither position.
The circle indicates the photometric aperture with a radius of $2\lambda/D$.
}
\label{fig:stackimage}
\end{figure*}

To investigate a possible mitigation of this effect, we simulated a $3\times3$ dither grid with a spacing of $1\lambda/D$. The dithered frames were combined in detector coordinates without image registration, producing a broadly illuminated ``sweet spot'' while preserving the spatial detector-response pattern, as illustrated in Figure~\ref{fig:stackimage}. A relative positioning error of $0.1\lambda/D$ was applied independently to each dither position.

We performed aperture photometry using a circular aperture with a radius of $2\lambda/D$. The aperture encloses approximately $70\%$ of the total flux in the combined dithered images. Assuming photon-noise-limited observations, the effective signal-to-noise ratio within the aperture was calculated as
\begin{equation}
{\rm SNR}_{\rm ap}
=
\sqrt{f_{\rm enc}}\,
{\rm SNR}_{\rm total},
\end{equation}
where $f_{\rm enc}=0.70$ is the enclosed-flux fraction. The resulting effective SNR was used in the Monte Carlo estimate of the achievable LPF accuracy.

\subsection{Current Best Estimates}

We calculated the current best estimates of the LPF reconstruction accuracy in Bands~1 and 4 using the calibrators listed in Tables~\ref{table_b1} and \ref{table_b4}, respectively. The simulations adopted the same detector-response and photometric-noise assumptions as Zellem et al. (2022)\cite{2022SPIE12180E..1ZZ}, together with the dithered observing configuration described above.
For each calibration scenario, the nominal telescope-roll orientation was selected so that the instrument-frame PAs of the 2 polarized calibrators were approximately aligned with the instrumental $Q$ and $U$ axes. The observations were then simulated at roll offsets of $\pm15^\circ$ from this nominal orientation. This favorable geometry improves the conditioning of the Mueller-matrix solution.

Table~\ref{table_b1} summarizes the current Band~1 estimates for the 2 calibration scenarios and compares them with the estimate of Zellem et al. (2022)\cite{2022SPIE12180E..1ZZ}. The estimated LPF reconstruction errors are comparable to the previous estimate.
A small nonzero LPF bias remains because the weakly polarized standard is treated as unpolarized in the Mueller-matrix reconstruction, while its measured nonzero polarization is included when generating the simulated observations. In both scenarios, this bias is smaller than the scatter in the reconstructed LPF.
Table~\ref{table_b4} presents the corresponding current estimates for Band~4. The reconstruction errors are slightly larger than those for Band~1, primarily because the polarization fractions of the selected calibrators decrease toward longer wavelengths, as expected from the Serkowski law.

\begin{table}[htbp]
\centering
\small
\caption{Current best estimates of the achievable LPF accuracy in Band 1 for two calibration scenarios: coronagraphic observations for exoplanet and disk science, and calibration using bright reference stars observed through an ND filter.}
\label{table_b1}
\begin{tabular}{lcccccc}
\hline
Error Term & CBE & LPF rms & CBE & LPF rms & CBE & LPF rms \\
 & Zellem+22 & Zellem+22 & FPM & FPM & ND & ND \\
\hline
Calibrator LPF [\%] & 0 $\pm$ 0 &  & 0.10 $\pm$ 0.02 & 0.31\% & 0.10 $\pm$ 0.04 & 0.48\% \\
 & 2.7 $\pm$ 0.1 & 0.36\% & 4.15 $\pm$ 0.20 & 0.99\% & 2.44 $\pm$ 0.06 & 0.65\% \\
 & 4.3 $\pm$ 0.1 &  & 6.73 $\pm$ 0.13 & 0.50\% & 8.08 $\pm$ 0.05 & 0.49\% \\
Calibrator PA [deg] & 30 $\pm$ 0 &  & 78.70 $\pm$ 6.50 & 0.31\% & 105.36 $\pm$ 10.32 & 0.47\% \\
 & 91 $\pm$ 1 & 0.36\% & 64.90 $\pm$ 0.20 & 0.32\% & 116.76 $\pm$ 0.60 & 0.52\% \\
 & 32 $\pm$ 1 &  & 126.93 $\pm$ 0.90 & 0.46\% & 131.24 $\pm$ 0.69 & 0.53\% \\
Flat fielding error [\%] & 0.50 & 0.35\% & 0.50 & 0.51\% & 0.50 & 0.62\% \\
CTI error [\%] & 1.0 & 0.71\% & 1.00 & 0.88\% & 1.00 & 0.94\% \\
Phot. noise on target [\%] & 1.00 & 0.51\% & 1.00 & 0.65\% & 1.00 & 0.74\% \\
Phot. noise on calibrators [\%] & 0.10 & 0.70\% & 0.33 & 0.60\% & 0.33 & 0.88\% \\
 & 0.10 & 0.58\% & 0.33 & 0.52\% & 0.33 & 0.84\% \\
 & 0.10 & 0.58\% & 0.33 & 0.41\% & 0.33 & 0.52\% \\
\hline
Total LPF rms &  & 1.66\% &  & 1.81\% &  & 1.74\% \\
Total LPF offset &  & +0.00\% &  & -0.31\% &  & -0.43\% \\
\hline
\end{tabular}
\end{table}
\begin{table}[htbp]
\centering
\small
\caption{Same as Table~\ref{table_b1}, but for Band~4.}
\label{table_b4}
\begin{tabular}{lcccc}
\hline
Error Term & CBE & LPF rms & CBE & LPF rms \\
 & FPM & FPM & ND & ND \\
\hline
Calibrator LPF [\%] & 0.08 $\pm$ 0.02 & 0.32\% & 0.09 $\pm$ 0.03 & 0.45\% \\
 & 3.11 $\pm$ 0.21 & 1.37\% & 2.29 $\pm$ 0.07 & 0.76\% \\
 & 6.29 $\pm$ 0.17 & 0.65\% & 6.94 $\pm$ 0.09 & 0.52\% \\
Calibrator PA [deg] & 78.70 $\pm$ 6.50 & 0.32\% & 105.36 $\pm$ 10.32 & 0.43\% \\
 & 64.90 $\pm$ 0.20 & 0.33\% & 116.76 $\pm$ 0.60 & 0.48\% \\
 & 126.93 $\pm$ 0.90 & 0.47\% & 131.24 $\pm$ 0.69 & 0.51\% \\
Flat fielding error [\%] & 0.50 & 0.54\% & 0.50 & 0.61\% \\
CTI error [\%] & 1.00 & 0.93\% & 1.00 & 0.97\% \\
Phot. noise on target [\%] & 1.00 & 0.69\% & 1.00 & 0.74\% \\
Phot. noise on calibrators [\%] & 0.33 & 0.73\% & 0.33 & 0.92\% \\
 & 0.33 & 0.65\% & 0.33 & 0.87\% \\
 & 0.33 & 0.44\% & 0.33 & 0.52\% \\
\hline
Total LPF rms &  & 2.25\% &  & 1.92\% \\
Total LPF offset &  & -0.28\% &  & -0.39\% \\
\hline
\end{tabular}
\end{table}

\section{Progress on the End-to-End Test and Future Work}

Real observations are affected by several factors that are not included in the simplified Monte Carlo simulation described above, including dark current, image distortion, cosmic rays, and spatial variations in detector response. These effects can influence both the measured photometry of the calibration standards and the derived Mueller-matrix elements. End-to-end simulations are therefore required to assess the feasibility and achievable accuracy of the on-sky calibration approach considered here.

Through the Roman Coronagraph Community Participation Program (CPP; Savransky et al., 2024)\cite{2024SPIE13092E..1IS}, several software packages have been developed and made publicly available on GitHub, most notably \texttt{corgiDRP} and \texttt{corgisim} (Millar-Blanchaer et al., 2024\cite{2024SPIE13092E..56M}, Wang et al., 2026\cite{wang2026}, Zhang et al., 2026\cite{zhang2026}). \texttt{corgiDRP} provides functions for processing both science and calibration data, including the derivation of Mueller-matrix coefficients, while \texttt{corgisim} generates simulated Roman Coronagraph datasets.

\begin{figure*}[htbp]
\begin{center}
\includegraphics[width=\linewidth]{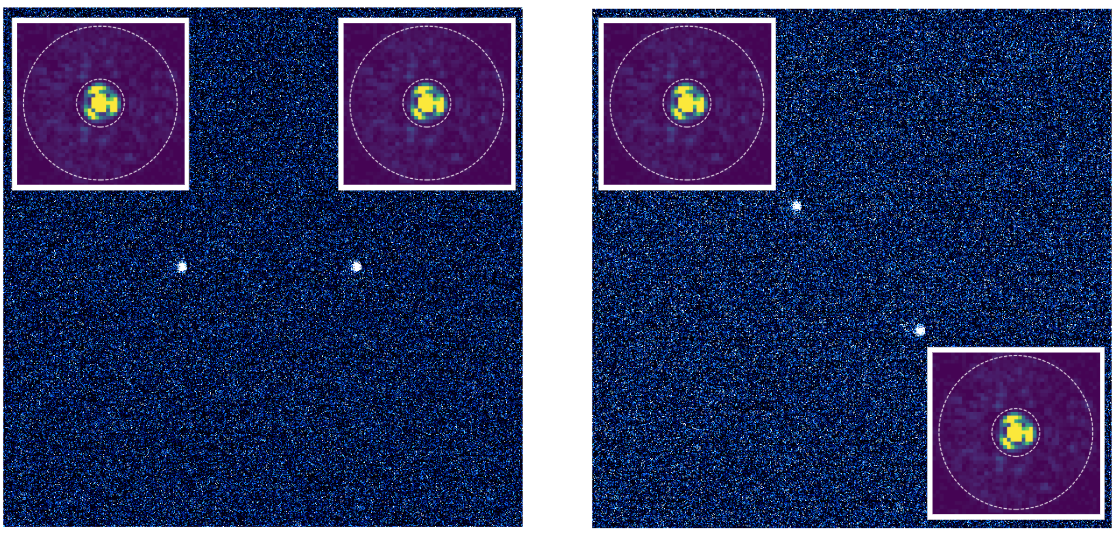}
\end{center}
\caption{
Four orthogonal polarization images, $I_0$, $I_{90}$, $I_{45}$, and $I_{135}$, of a polarized standard star simulated with \texttt{corgisim} and processed with \texttt{corgiDRP}.
}
\label{fig:simimage}
\end{figure*}

A complete end-to-end test of the polarimetric calibration mode is currently under development, and the realistic input parameters and observing configurations for these simulations have not yet been fully established. Using the latest version of \texttt{corgisim}, we generated simulated Level~1 datasets for a coronagraphic calibration scenario. These datasets were successfully processed from Level~1 through Level~2b. Figure~\ref{fig:simimage} shows an example of the resulting Level~2b images.

We are currently refining several pipeline step functions, particularly those required to incorporate the dithered observing strategy into the Mueller-matrix derivation. We are also updating the calibrator information and testing the subsequent processing steps required to derive the Mueller-matrix coefficients.

\section{Summary}

We developed a calibration framework for polarimetric observations with the \cgi\ using on-sky polarized and weakly polarized standard stars. Candidate calibrators were selected by combining published optical polarimetry with Gaia DR3 astrometry and photometry and applying constraints on ecliptic latitude, brightness, nearby sources, astrometric quality, and polarization fraction. Separate bright and faint samples were considered for calibration observations using an ND filter and for coronagraphic calibration observations without an ND filter, respectively.

Precursor $VRI$-band observations of 18 faint candidates were obtained with HONIR on the 1.5-m Kanata telescope. The measurements were combined with published polarimetry and fitted in Stokes space using the Serkowski law. These fits were used to predict the polarization properties of the candidates in \cgi\ Bands~1 and 4. Two sets of 3 calibrators, each consisting of 1 weakly polarized and 2 polarized standards, were selected for the 2 calibration scenarios.

The achievable LPF reconstruction accuracy was evaluated using Monte Carlo simulations of the Mueller-matrix calibration. The simulations distinguish between the nominal calibrator Stokes vectors adopted in the calibration and the perturbed Stokes vectors used to generate simulated observations. Uncertainties in the calibrator polarization properties, residual detector-response errors, and photometric noise were included. We also investigated a $3\times3$ dithered observing configuration to increase the detector area commonly sampled by the calibrators and thereby reduce differential detector-response errors. The current estimates indicate LPF reconstruction errors at the few-percentage-point level in Bands~1 and 4. A small residual bias remains when the measured nonzero polarization of the weakly polarized standard is neglected in the Mueller-matrix reconstruction.

A complete end-to-end validation of the calibration procedure is still in progress. Simulated \cgi\ datasets generated with \texttt{corgisim} have been successfully processed from Level~1 through Level~2b using \texttt{corgiDRP}. Ongoing work includes incorporating the dithered observing configuration into the Mueller-matrix derivation, updating the adopted calibrator information, and testing the remaining processing steps required to evaluate the polarimetric calibration accuracy under more realistic observing and detector conditions.

\acknowledgments 
This research was carried out in part at the Jet Propulsion Laboratory, California Institute of Technology, under a contract with the National Aeronautics and Space Administration (80NM0018D0004).
J.\ Hom is supported by NASA under award No. 80NSSC25K0364.
Portions of this work were supported by the NASA under Grant 80NSSC25K0367 issued through the Roman Community Participation program (PI, Anche).
M.A.M.-B.\ was supported by NASA under award 80NSSC24K0097.
J.J.\ Wang was supported by NASA under award 80NSSC24K0087 and by the Sloan Foundation.
This material is based upon work by S.G.W. supported by NASA under award No. 80NSSC24K0217.
\bibliography{report} 
\bibliographystyle{spiebib} 

\end{document}